\documentclass[fleqn,usenatbib]{class}

\usepackage{newtxtext,newtxmath}

\usepackage{float}
\usepackage{wrapfig}
\usepackage{upgreek}
\restylefloat{figure}
\restylefloat{table}
\usepackage{tabularx}

\makeatletter
\newcommand{\thickhline}{%
    \noalign {\ifnum 0=`}\fi \hrule height 2pt
    \futurelet \reserved@a \@xhline
}

\newcolumntype{"}{@{\hskip\tabcolsep\vrule width 2pt\hskip\tabcolsep}}    
\makeatother

\usepackage[paper=a4paper,left=17mm,right=17mm,top=20mm,bottom=20mm]{geometry}

\usepackage{caption}
\usepackage{setspace}

\usepackage{tabu}

\usepackage{natbib}

\usepackage{color,soul}

\usepackage{enumitem}
\usepackage{ulem}
\usepackage{xcolor,colortbl}

\definecolor{customdeepgreen}{HTML}{006A71}

\makeatletter
    \newcommand{\coloruline}[2]{%
        \newcommand\temp@reduline{\bgroup\markoverwith
            {\textcolor{#1}{\rule[-0.5ex]{2pt}{0.4pt}}}\ULon}%
        \temp@reduline{#2}%
    }

    \newcommand{\coloruuline}[2]{%
        \UL@protected\def\temp@uuline{\leavevmode \bgroup
            \UL@setULdepth
            \ifx\UL@on\UL@onin \advance\ULdepth2.8\p@\fi
            \markoverwith{\textcolor{#1}{\lower\ULdepth\hbox
                {\kern-.03em\vbox{\hrule width.2em\kern1\p@\hrule}\kern-.03em}}}%
        \ULon}%
        \temp@uuline{#2}%
    }

    \newcommand{\coloruwave}[2]{%
        \UL@protected\def\temp@uwave{\leavevmode \bgroup 
        \ifdim \ULdepth=\maxdimen \ULdepth 3.5\p@
        \else \advance\ULdepth2\p@ 
        \fi \markoverwith{\textcolor{#1}{\lower\ULdepth\hbox{\sixly \char58}}}\ULon}
        \font\sixly=lasy6 % does not re-load if already loaded, so no memory drain.
        \temp@uwave{#2}%
    }
\makeatother

\usepackage{hyperref}
\hypersetup{
    colorlinks=true,
    linkcolor=ultramarine3,
    filecolor=ultramarine3,      
    urlcolor=ultramarine3,
    pdftitle=T(z) and UHECR propagation,
    pdfpagemode=FullScreen,
    citecolor=ultramarine3
    }

\definecolor{ultramarine}{RGB}{0,32,96}
\definecolor{ultramarine2}{RGB}{0,68,204}
\definecolor{ultramarine3}{RGB}{0,0,180}

\usepackage[T1]{fontenc}

\DeclareRobustCommand{\VAN}[3]{#2}
\let\VANthebibliography\thebibliography
\def\thebibliography{\DeclareRobustCommand{\VAN}[3]{##3}\VANthebibliography}

\usepackage{graphicx,subfigure}	% Including figure files
\usepackage{amsmath}	% Advanced maths commands
\usepackage{amsmath,array,graphicx}
\usepackage{kantlipsum}
\usepackage{hyperref}
\usepackage{verbatim}

\newcommand{\eqb}{\begin{equation}}
\newcommand{\eqe}{\end{equation}}
\newcommand{\dmb}{\begin{displaymath}}
\newcommand{\dme}{\end{displaymath}}

\newcommand{\eab}{\begin{eqnarray}}
\newcommand{\eae}{\end{eqnarray}}

\newcommand{\be}{\begin{equation}}
\newcommand{\ee}{\end{equation}}

\title[$T(z)$ and UHECR propagation]
{Modified temperature redshift relation and UHECR propagation}

\author[Meinert et al.]{
\noindent Janning Meinert$^{1,2}$\thanks{meinert@uni-wuppertal.de (corresponding author)}\href{https://orcid.org/0000-0001-7582-3456}{\hspace{0.1mm}\includegraphics[scale=0.06]{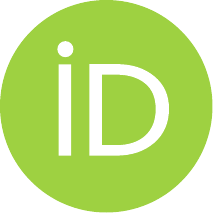}}, 
Leonel Morejón$^{1}$\thanks{leonel.morejon@uni-wuppertal.de}\href{https://orcid.org/0000-0003-1494-2624}{\hspace{0.1mm}\includegraphics[scale=0.06]{orcid.pdf}},\vspace{2mm}
Alexander Sandrock$^{1}$\thanks{asandrock@uni-wuppertal.de}\href{https://orcid.org/0000-0002-6779-1172}{\hspace{0.1mm}\includegraphics[scale=0.06]{orcid.pdf}},\vspace{2mm} 
Björn Eichmann$^{3}$\thanks{eiche@tp4.rub.de},\vspace{2mm}
\\
\vspace{-8mm}
\newauthor\noindent\hspace{-0mm} 
Jonas 
Kreidelmeyer$^{4}$\thanks{jonas.kreidelmeyer@desy.de}\href{https://orcid.org/0000-0002-6185-6414}{\hspace{0.1mm}\includegraphics[scale=0.06]{orcid.pdf}},\vspace{2mm} 
Karl-Heinz Kampert$^{1}$\thanks{kampert@uni-wuppertal.de}\href{https://orcid.org/0000-0002-2805-0195}{\hspace{0.1mm}\includegraphics[scale=0.06]{orcid.pdf}}\vspace{2mm} 
\vspace{3mm}
\\
$^{1}$Bergische Universit\"{a}t Wuppertal, Department of Physics,
  Gau\ss stra\ss e 20, 42103 Wuppertal, Germany\vspace{1mm}
\\
$^{2}$Institut f\"ur Theoretische Physik, Universit\"at Heidelberg Philosophenweg 12, D-69120 Heidelberg, Germany\vspace{1mm}
\\
$^{3}$Ruhr-Universität Bochum, Theoretische Physik IV, Fakultät für Physik und Astronomie, Universitätsstraße 150, 44801 Bochum, Germany\vspace{1mm}
\\
$^{4}$Institut für Experimentalphysik, Universität Hamburg, DESY, Luruper Chaussee 149, Hamburg, Germany\vspace{1mm}
}

\begin{document}
\label{firstpage}
\pagerange{\pageref{firstpage}--\pageref{lastpage}}
\maketitle

\begin{abstract}
We re-examine the interactions of ultra-high energy cosmic rays (UHECRs) with photons from the cosmic microwave background (CMB) under a changed, locally non-linear temperature redshift relation $T(z)$. This changed temperature redshift relation has recently been suggested by the postulate of subjecting thermalised and isotropic photon gases such as the CMB to an SU(2) rather than a U(1) gauge group. This modification of $\Lambda$CDM is called SU(2)$_{\rm CMB}$, and some cosmological parameters obtained by SU(2)$_{\rm CMB}$ seem to be in better agreement with local measurements of the same quantities, in particular $H_0$ and S$_8$.
In this work, we apply the reduced CMB photon density under SU(2)$_{\rm CMB}$ to the propagation of UHECRs. This leads to a higher UHECR flux just below the ankle in the cosmic ray spectrum and slightly more cosmogenic neutrinos under otherwise equal conditions for emission and propagation. Most prominently, the proton flux is significantly increased below the ankle ($5\times10^{18}$\,eV) for hard injection spectra and without considering the effects of magnetic fields. 
The reduction in CMB photon density also favours a decreased cosmic ray source evolution than the best fit using $\Lambda$CDM. 
In consequence, it seems that SU(2)$_{\rm CMB}$ favours sources that evolve as the star formation rate (SFR),
like starburst galaxies (SBG) and gamma-ray bursts (GRB), over active galactic nuclei (AGNs) as origins of UHECRs.
We conclude that the question about the nature of primary sources of UHECRs is directly affected by the assumed temperature redshift relation of the CMB.
\end{abstract}


%%%%%%%%%%%%%%%%%%%%%%%%%%%%%%%%%%%%%%%%%%%%%%%%%%

%%%%%%%%%%%%%%%%% BODY OF PAPER %%%%%%%%%%%%%%%%%%

\section{Introduction}

The cosmic microwave background (CMB) is the cornerstone of modern Cosmology. Modelling its properties correctly is, however, not only relevant for Cosmology but also vital for the correct description of ultra-high energy cosmic ray (UHECRs) propagation. 

In this work, we dilute the CMB photon density in comparison to the standard cosmological model $\Lambda$CDM, by assuming the so-called SU(2)$_{\rm CMB}$ model \citep{Hofmann:2009yh,Hahn:2018dih,bookHofmann,Hofmann:2022qsa,Hofmann:2023wyk}.
The purpose of this paper is to discuss how these potential changes to the CMB photon density influence the propagation
of UHECRs.\\

Previous discussions of the consequences of an SU(2)$_{\rm CMB}$ description on UHECR interactions were limited to considering the handedness of the photons, SU(2)$_{\rm L}$ \citep{Tipler:2018iqo}.
A fully consistent understanding of the SU(2)$_{\rm CMB}$ model requires applying Yang-Mills thermodynamics and obtaining the modified $T(z)$. Furthermore, the effect of this modified temperature redshift relation on the CMB density produces non-trivial redshift dependences on the UHECR interactions that need to be considered in depth. 
For a discussion of the impact of modified gravity on UHECR propagation, please see \citep{Sarmah:2023dod}.
Firstly, the modified $T(z)$ relation is outlined in section \ref{TZsu2}. The consequences of this relation for all the interactions of UHECRs are discussed in section \ref{UHECRs}. Section \ref{Model Comparison} compares fits of UHECRs spectral energy and composition measured by the Pierre Auger Observatory with both U(1) and SU(2) $T(z)$ relations. The corresponding cosmogenic neutrino fluxes are presented in section \ref{Cosmogenic Neutrinos Model Comparisonn}. 

\vspace{-4mm}

%####################################################
%                  Changed T(z)
%####################################################
\section{
%\texorpdfstring{$T(z)$}{{\textit{\textbf{ T(z)}}}}
{\textit{\textbf{ T({\scriptsize z})}}} relation of \texorpdfstring{SU(2)$_{\rm CMB}$}{SU(2)CMB}}\label{TZsu2}
%###################################

In the following, we briefly review the $T(z)$ relation of deconfining SU(2)$_{\rm CMB}$ thermodynamics. For a longer version of the argument, the reader is referred to \cite{Hahn:2017yei,Hofmann:2023wyk}. The core idea is that the additional degrees of freedom in an SU(2) gauge group lead to the topological constant $1/4^{1/3}$, so that the $T(z)$ relation is for $z \gg 1$ given by
\eqb
\label{T-z-relation}
T(z)/T_0 =\left(\frac14\right)^{1/3}(1 + z)\,,\ \ \ \ \ (T(z) \gg T(z=0)). 
\eqe

\begin{figure}[tb]
\centering
\includegraphics[width=\columnwidth]
{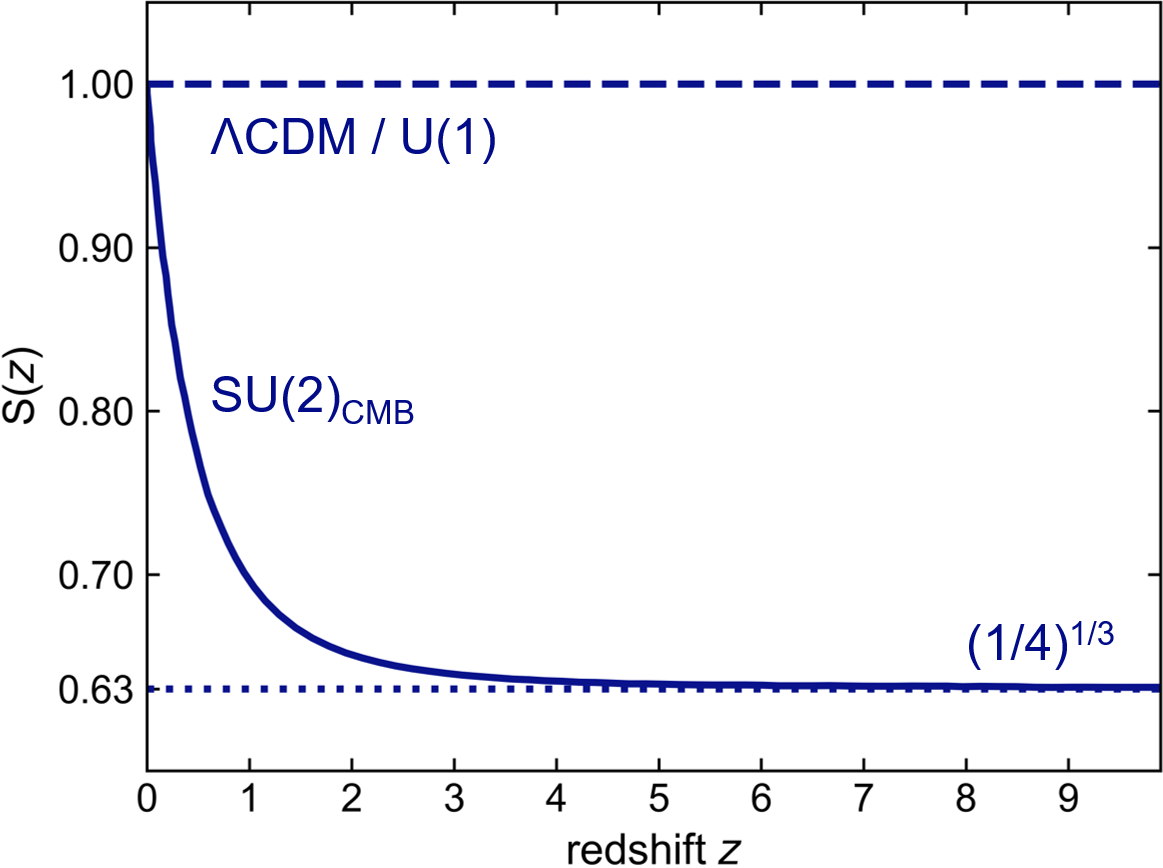}%{SU2TemperatureRedshiftRelation.png}
\caption{\protect{\label{fig:Sz}} Plot of function ${\cal S}(z)$ in Eq.\,(\ref{soltT0}) for SU(2)$_{\rm CMB}$ in solid.
The conventional $T(z)$ relation of the CMB, as used in the cosmological standard model $\Lambda$CDM, associates with the %blue 
dashed line ${\cal S}(z)\equiv 1$.
The high-temperature value $1/4^{1/3}$ is approximated by the dotted line  ${\cal S}(z) = 0.63$.
}
\end{figure}

To derive this constant, a flat Friedmann-Lema\^{i}tre-Robertson-Walker (FLRW) universe is assumed:
\eqb
\label{enecons}
\frac{\mbox{d}\rho}{\mbox{d}a}=-\frac{3}{a}\left(\rho+P\right)\,,
\eqe
where $\rho$ denotes the energy density, and $P$ the pressure of the deconfined phase in SU(2) thermodynamics. The scale factor $a$ is dimensionless, $a(T(z=0))=1$, and related to the redshift $z$ according to $1/a=z+1$. Eq.\,(\ref{enecons}) has the solution 
\eab
\label{formsol}
%\begin{split}
a &=&\exp\left(-\frac{1}{3}\int^{\rho(T)}_{\rho(T(z=0))}\frac{\mbox{d}\rho}{\rho+P(\rho)}\right) \nonumber\\
&=& \exp\Biggl(-\frac{1}{3}\int^T_{T(z=0)}
\, \underbrace{\frac{1}{T^\prime} \frac{\mbox{d}\rho}{\mbox{d}T^\prime}}_{\kappa}
\,\frac{\mbox{d}T^\prime}{s(T^\prime)}\Biggr) \,,
\eae
where the entropy 
density $s$ is defined as $s=(\rho+P)/{T}$. By using the Legendre transformation 
\eqb
\label{LT}
\rho=T\frac{\mbox{d}P}{\mbox{d}T}-P\,,
\eqe
the term $\kappa$ can be expressed as
\eqb
\label{Tdiff}
\kappa = \frac{1}{T}\frac{\mbox{d}\rho}{\mbox{d}T}=
\frac{\mbox{d}^2P}{\mbox{d}T^2}=
\frac{\mbox{d}s}{\mbox{d}T}\,.
\eqe 
Substituting Eq.\,(\ref{Tdiff}) into Eq.\,(\ref{formsol}) finally yields
\eqb
\label{sol}
a 
= \exp\left(-\frac13\log\frac{s(T)}{s(T(z=0))}\right)\,.
\eqe 
The formal solution (\ref{sol}) is valid for any thermal and conserved 
fluid subject to expansion in an FLRW universe. 
If the function $s(T)$ is known, then $T(z)$ can be derived. 
The ground-state of the deconfining phase is independent of the $T(z)$ relation, since the equation of state for ground-state pressure $P^{\rm gs}$ and energy density $\rho^{\rm gs}$ is $P^{\rm gs}=-\rho^{\rm gs}$ \citep[see also][]{bookHofmann}. 
Asymptotic freedom occurs nonperturbatively for $T(z)\gg T(z=0)$ \citep{Gross:1973id,Politzer:1973fx,bookHofmann}, and therefore $s(T)$ is proportional to $T^3$. Due to a decoupling of massive vector modes at $T(z=0)$, excitations represent a free photon gas. Therefore, $s(T(z=0))$ is also proportional to $T^3(z=0)$. Correspondingly, the ratio $s(T)/s(T(z=0))$ in Eq.\,(\ref{sol}) reads
\eab
\label{ratentr}
%\begin{split}
\frac{s(T)}{s(T(z=0))} 
=\frac{g(T)}{g(T(z=0))} \left(\frac{T}{T(z=0)}\right)^3\,,\,(T\gg T(z=0))\,, 
\eae
where $g$ refers to the number of relativistic degrees of freedom at the respective 
temperatures. SU(2) has one massless gauge mode with two polarisations and two massive gauge modes with three polarisations each, so $g(T)=2\times 1+3\times 2=8$, for U(1) there is only one massless mode, $g(T(z=0))=2\times 1$. Substituting this into Eq.\,(\ref{ratentr}), inserting the result into Eq.\,(\ref{sol}), and solving for $T$, one arrives at the high-temperature $T(z)$ relation 
\eab
\label{solt>t0}
T(z)&=&\left(\frac14\right)^{1/3}(z+1)\,T(z=0)\,,\ \ \  (T\gg T(z=0))\,. %\nonumber\\ 
%&\approx&0.629\,\,(z+1)\,T(z=0)\,,\ \ \  (T\gg T(z=0))\,.
\eae
Due to two massive vector modes contributing to $s(T)$ at low temperatures, the $T(z)$ relation is modified to 
\begin{equation}
\label{soltT0}
T(z) = {\cal S}(z)(z+1)\,T(z=0)\,, 
\ \ \  (T\ge T(z=0))\,,
\end{equation}
where the nonlinear function ${\cal S}(z)$ is depicted in Fig.\,\ref{fig:Sz} and derived in \cite{Hahn:2018dih}. 
The function ${\cal S}(z)$ can be approximated reasonably well with the analytical function
\eqb
\label{SZapprox}
{\cal S}(z)_{\rm SU(2)} \approx {\rm exp}(-1-1.7\,z)+\left(\frac14\right)^{1/3}\,.
\eqe 
%\vspace{-0.1mm}
This approximation will be used in section \ref{UHECRs}. However, the numerical solution was applied for all following sections.

%***********************
%\vspace{-1cm}
%************************

%####################################################
%            Changes in propagation length
%####################################################
%\newpage
%\vspace{-1cm}

\section{Changes in propagation length}\label{UHECRs}

In this section, we discuss the changes  to the propagation of ultra-high energy cosmic rays produced by employing the modified temperature relation $T(z)$ from SU(2)$_{\rm CMB}$ as derived in the previous section, Eqs.\,\ref{solt>t0} and \ref{soltT0}.

\begin{figure}[ht]%[tbh]
\centering
\caption{\protect{\label{propagationProton}} Propagation length of protons at redshift $z = 1$ as a function of the initial particle energy. The normal U(1) and the SU(2) induced $T^{\prime}(z)$ propagation lengths are shown as dashed and full lines, respectively.}
\includegraphics[width=\columnwidth]{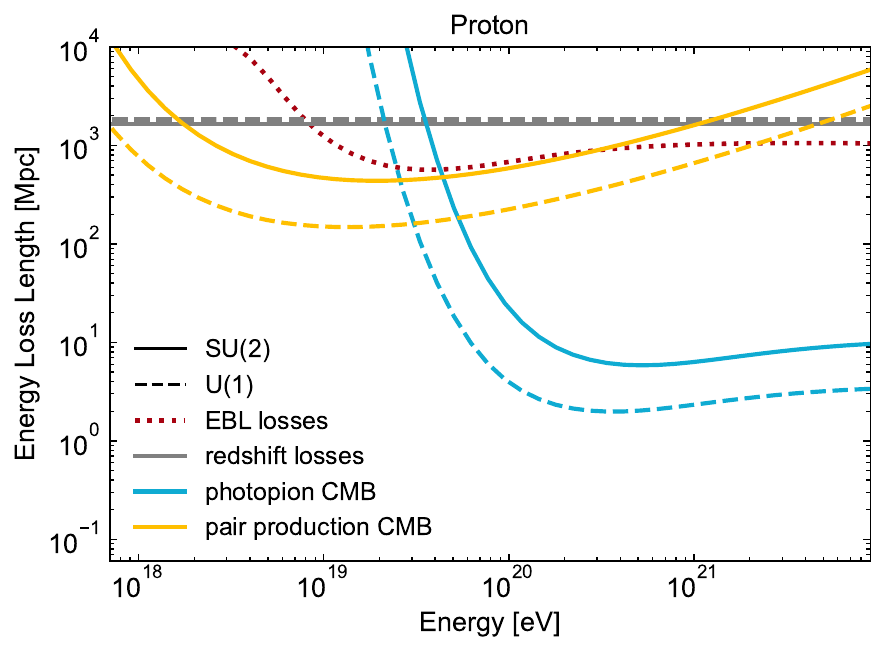} 
\end{figure}

\begin{figure}[h]%[tbh]
\centering
\caption{\protect{\label{propagationIron}} 
As Fig.\,\ref{propagationProton} for the propagation length of iron nuclei.
}
\includegraphics[width=\columnwidth]{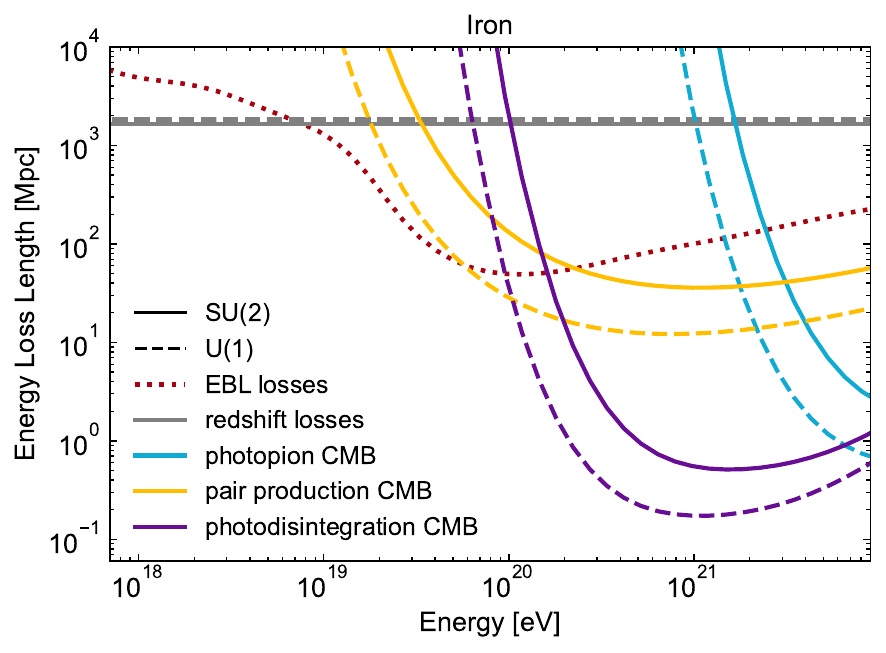} 
\end{figure}

The redshift dependence of the CMB temperature results in scaling and shifting of the differential CMB photon number density $n_{\rm CMB}(\epsilon, z)$ 
\begin{equation}
    \label{eq:density_scaling}
    n_{\rm CMB}(\epsilon, z) = \left(\frac{T(z)}{T_0}\right)^2\,
   n_{\rm CMB}\left(\epsilon \left(\frac{T(z)}{T_0}\right)^{-1}, 0\right)\,,
\end{equation}
where $\epsilon$ is the energy of the photons, and $n_{\rm CMB}$ as derived from the Planck distribution is
\begin{equation}
    n_{\rm CMB}(\epsilon, z)\,= 
    %(c\hbar)^{-3} \left(\frac{\epsilon}{\pi}\right)^2 %\left(\exp(\epsilon/k_B T(z)) - 1\right)^{-1}
    \frac{1}{\pi^2 c^{3} \hbar^{3}}
   \frac{\epsilon^2}{\exp(\epsilon/k_B T(z)) - 1}
\end{equation}
where $k_B$ is the Boltzmann constant.
The redshift dependence of UHECR interactions with the CMB is reflected in the expression for the energy loss length \citep{Berezinskji1990}

\begin{equation*}
    \label{eq:ELL}
    -\frac{1}{E}\frac{dE}{dx} = \int_{\epsilon_0}^{\infty}
    \frac{k_BT\, d\epsilon'\sigma(\epsilon')f(\epsilon')\epsilon'}{2 \pi^2\Gamma^2 c^3\hbar^3}\,
    \left\{-\ln \left[ 1 - \exp \left( -\frac{\epsilon'}{2\Gamma k_BT}\right)\right]\right\}
\end{equation*} 

where $E$ is the energy and $\Gamma=(1-(v/c)^2)^{-1/2}$ is the Lorentz boost of the UHECRs, and $\sigma(\epsilon')$ is the cross-section for the corresponding interaction (photodisintegration, photomeson, pair-production) and $f(\epsilon')$ is the average inelasticity of the interaction. The scaling of the CMB density produces a corresponding scaling of the interaction rates $\lambda(\Gamma, z)$:
\begin{equation}
    \lambda(\Gamma, z) = \left(\frac{T(z)}{T_0}\right)^3\, \lambda \,\left(\frac{T(z)}{T_0}\Gamma, z=0 \right)\,.
    \label{eq:rates_scaling}
\end{equation}
The comparison of the energy loss lengths for U(1) and SU(2) is shown in Fig.\,\ref{propagationProton} (protons) and in Fig.\,\ref{propagationIron} (iron) for $z=1$. The interaction processes with the CMB are represented separately (photopion, photodisintegration, pair production) while they are grouped into one curve for extragalactic background light (EBL, dotted dark red)\footnote{In this work we do not include the CMB nor the radio background into the description of the EBL. The EBL remains unchanged under the assumption of an SU(2) gauge group for thermal photons, as the EBL is not thermalized.}. For protons at redshift $z=1$ the energy loss length at the GZK-limit ($E\sim$\,5$\,\times$\,$10^{19}$\,eV) is shifted by a factor of  $\sim$2 to higher energies for SU(2) and the propagation lengths for both pair production and photopion production are increased by nearly a factor 3. For iron nuclei at the same redshift, the corresponding photodisintegration limit is also shifted to higher energies by a factor $\sim$2 for the SU(2). However, because the energy loss lengths are also increased due to the reduced CMB density, the interactions with the EBL are the dominant ones for cosmic ray energies below $10^{20}$\,eV and therefore the total energy loss length is not increased as much as in the case of protons. 
This is representative of the case for all intermediate nuclear species with masses between the proton and iron. 
The increase in energy loss lengths implies the expansion of the horizon for UHECRs: for protons at all energies, for nuclei at the highest energies starting from about $\sim10^{19}$\,eV. With such an increase, protons from sources at redshift 1 and energies $(1-40)\,\times\,10^{19}$\,eV would propagate for several hundreds of megaparsecs more than in the case of the U(1), whereas protons at higher energies (where the photopion interactions prevail) would propagate for more than ten megaparsecs.\\

These increases of propagation horizons are only important when the contribution from distant sources is the dominant one. As the redshift evolves to the present, the U(1) and SU(2)$_{\rm CMB}$ densities converge and by distances of 20\,Mpc from Earth the loss lengths differ by only 1.5\,\%. Thus, although protons can propagate further away from sources beyond $\sim$200\,Mpc in the SU(2) case, they completely lose their energy before reaching our galaxy and only the secondary neutrinos reach us, much like in the U(1) case.

%\newpage
\vspace{-1mm}

Nonetheless, protons coming from sources marginally closer are able to reach our galaxy: at a distance of 200\,Mpc Eq.\,\ref{eq:rates_scaling} yields a reduction in the interaction rates of $\sim$9\,\% for the SU(2) scenario, see Fig.\,\ref{propagationProton}.
For nuclei the increased propagation is, however, much less relevant since their propagation lengths are limited to a few dozens of Mpc. For such distances, the reduction in interaction rates with the CMB is of $2\,-\,4\,\%$ for SU(2). However, those interactions are overshadowed by the dominant interactions with the EBL.

%####################################################
%                   Heavy Nuclei
%####################################################

%\newpage

%\section{Propagation of UHECRs}
\section{Observational consequences for UHECR energy spectra}
\label{Model Comparison}

We evaluate the impact on the propagation of UHECRs by employing the fit obtained by \cite{Heinze:2019jou} to data from the Pierre Auger Observatory \citep{PierreAuger:2016use} under a conventional temperature redshift relation ($\Lambda$CDM). The changes in spectral energy and composition produced with the same fit values under SU(2)$_\textrm{CMB}$ are obtained by employing the modified $T(z)$-relation.
The propagation of UHECRs was performed using \href{https://github.com/joheinze/PriNCe}{PriNCe} \citep{Heinze:2019jou}, which is an efficient code to integrate the transport equations for the evolution of cosmic rays at cosmological scales. It includes all the relevant interactions and allows for custom modifications, however, it does not account for the effect of magnetic fields.
The propagation scenario considers a population of sources with a continuous distribution in redshift proportional to $(1+z)^m$ with source evolution parameter $m$ obtained from the fit. The sources are assumed to be isotropically distributed and to eject a rigidity-dependent spectral energy flux according to
\eqb
\label{CR-flux}
J_A (E) = \mathcal{J}_A\, f_{\rm cut}(E,Z_A,R_{\rm max})\, (1+z)^m \, \left(\frac{E}{E_0} \right)^{-\gamma},
\eqe 
\noindent
where $Z_A$ is the atomic number and the five nuclear mass groups are indicated by the index $A$ (denoting the nuclear species $^1$H, $^4$He, $^{14}$N, $^{28}$Si, and $^{56}$Fe). They share the same spectral index $\gamma$ and the maximal rigidity $R_{\rm max} = E_{\rm max} / Z_A$. The cutoff of the injection spectra $f_{\rm cut}$ is defined as 
\eqb
\label{fcut}
f_{\rm cut}(E) = \begin{cases}
      1, & E < Z_A R_{\rm max}\\
      {\rm exp}\left(1-E/(Z_A R_{\rm max})\right), & E>Z_A R_{\rm max}.
    \end{cases}
\eqe 
$\mathcal{J}_A$ represents the flux of particles of species $A$ emitted per unit of time, comoving volume, and energy. The elemental injection fractions $f_A$ are defined as $f_A = \mathcal{J}_A / (\Sigma_{A'} \, \mathcal{J}_{A'})$ at the reference energy $E_0 = 10^{18}$\,eV. 
Here, $\Sigma_{A'}$ denotes the sum over all chosen nuclear species.
Integrating over the injected fluxes $J_A$ leads to the integral fractions of the energy density $I_A$, which are independent of the choice of $E_0$:
\begin{align}
\label{I_A}
I_A= \frac{\int_{E_{\rm min}}^{\infty}  J_A E \,dE}{\Sigma_{A'} \int_{E_{\rm min}}^{\infty}  J_{A'} E \,dE}= \frac{\int_{E_{\rm min}}^{\infty}  f_A\,f_{\rm cut}(E,Z_A)\,E^{1-\gamma} dE}{\Sigma_{A'} \int_{E_{\rm min}}^{\infty}  f_{A'}\,f_{\rm cut}(E,Z_{A'})\,E^{1-\gamma} dE},
\end{align} 
\noindent where $E_{\rm min}=10^{18}$\,eV. For the sake of completeness, we provide both $f_A$ and $I_A$ in the following sections. 
For SU(2)$_\textrm{CMB}$ the following cosmological parameters were used for the propagation: The Hubble parameter $H_0$= 74.24 km\,s$^{-1}$Mpc$^{-1}$, a dark energy fraction of $\Omega_{\Lambda} = 0.616$, and the local matter density $\Omega_{\text{m,0}} = 0.384$, compare with \cite{Hahn:2018dih}.
For U(1)$_\textrm{CMB}$ ($\Lambda$CDM) the values from the Planck Collaboration were used 
\cite[p. 15, Table 2]{Planck:2018vyg}, where $H_0$ = 67.36 km\,s$^{-1}$Mpc$^{-1}$, $\Omega_{\Lambda} = 0.6847$ and $\Omega_{\text{m,0}} = 0.3153$ (TT,TE,EE+lowE+lensing).\\

\begin{figure}[tbh]
\centering
\includegraphics[width=\columnwidth]
%{pictures/Heinze.png}
{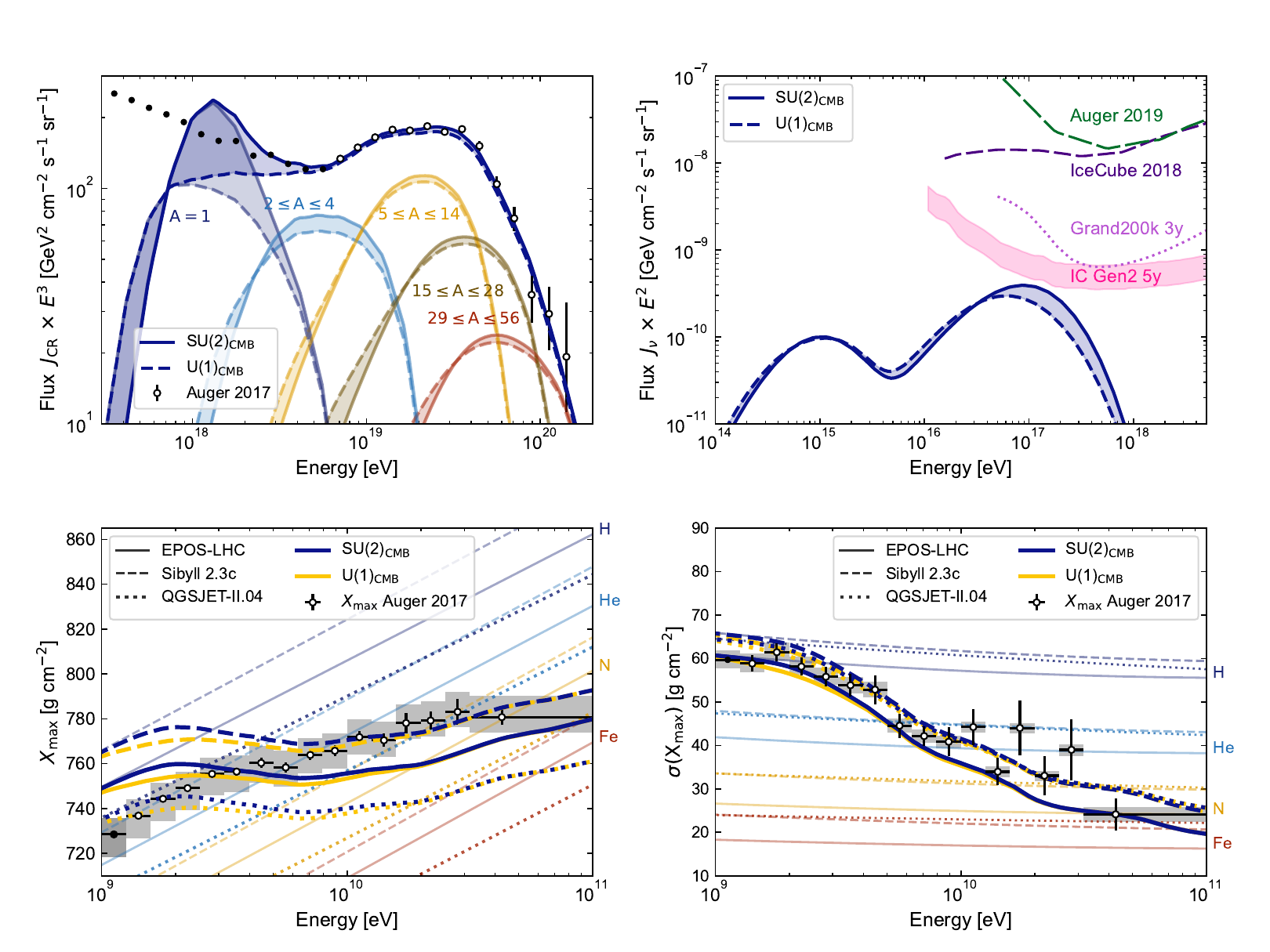}
\caption[]{
Spectral fit to the 2017 Auger spectral flux data \citep{PierreAuger:2016use}
from the best fit parameters in \protect\cite{Heinze:2019jou}, see Tab.\,\ref{table:HeinzeTable}. 
The fluxes of the normal U(1) and modified SU(2) temperature redshift relations are shown as dashed and full lines, respectively. The $\chi^2$ only considers data points above the ankle region (white dots), as was done in \cite{Heinze:2019jou}.
}
\label{fig:HeinzeBestFit}
\end{figure}

%{\color{red}
The best fit parameters obtained by \cite{Heinze:2019jou} for the conventional $\Lambda$CDM relation are reported in Table\,\ref{table:HeinzeTable} and plotted in Figure\,\ref{fig:HeinzeBestFit} for reference (dashed lines).  Fixing these source parameters and propagating the injected UHECR through the SU(2)$_{\rm CMB}$ with its modified $T(z)$ relation yields the solid lines in the same figure.
% I would not start a new paragraph here -- okay!
As can be seen, the resulting total flux for SU(2) is virtually unchanged for energies above $6\times10^{18}$\,eV, while the fluxes for individual nuclear groups show slightly more pronounced peaks. This effect is a consequence of the modest increase in the horizons. At the same time, the reduction in the pair production losses produces sharper peaks because the effect of energy redistribution corresponding to the U(1) cases is less prominent for SU(2). For protons at the lowest energies, the differences are much more pronounced due to the change in pair production rates as the energies approach $10^{18}$\,eV from above. 

\begin{table}[tbh]
\centering
\begin{tabu}{ | l | c |[1.5pt] c | c | c |  }
\hline
 EBL & Gilmore et al. & Element & $f_A$ \% & $I_A$ \% \\ \hline
 models & TALYS \& Sibyll 2.3c & H & 0.0 &   0.0\\  \hline 
 redshifts & $1\,-\,0$ & He & 82.0 & 9.91 \\  \hline
 $\gamma$ & $-0.8$ & Ni & 17.3 & 69.99 \\\hline
 $R_{\textrm{max}}$ & $1.6 \times 10^{18}$\,V & Si & 0.6 &16.91\\\hline
 $m$ & 4.2 & Fe & 0.02 & 3.19\\\hline
\end{tabu}
\protect\caption{Best fit parameters from \protect\cite{Heinze:2019jou}, Table 3.} %, p. 26.}
\label{table:HeinzeTable}
\end{table}

\begin{figure}[tbh]
\centering
\includegraphics[width=\columnwidth]
%{pictures/gradient_descent.png}
{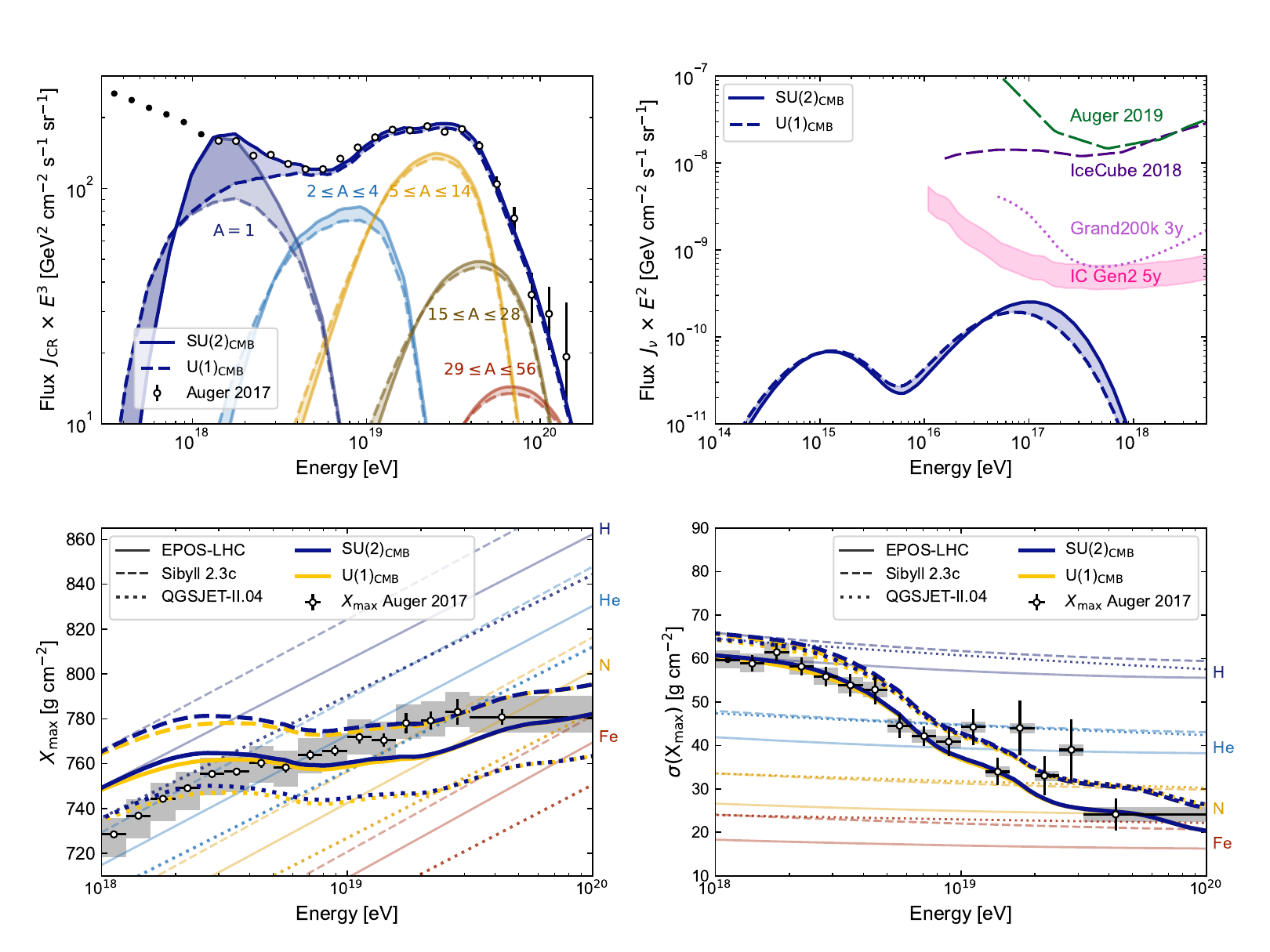}
\caption{
Spectral fit using a gradient descent algorithm to the 2017 Auger spectral flux data, $X_{\rm max}$ and $\sigma(X_{\rm max})$ data \citep{PierreAuger:2016use}.
Here, $X_{\rm max}$ denotes the position of the shower maximum in the atmosphere.
The fluxes of the normal U(1) and modified SU(2) temperature redshift relations are shown as dashed and full lines, respectively. 
%The total flux is shown for U(1) in a navy blue dashed line; for  SU(2) in a navy blue solid line.
The $\chi^2$ was computed including all the white dots.
}
\label{fig:HeinzeBestSU2}
\end{figure}
\newpage
\vspace{-1mm}

Repeating now the combined $E$, $X_{\rm{max}}$, and $\sigma(X_{\rm{max}})$ fit to the same data set of \cite{PierreAuger:2016use} employing a gradient descent algorithm \citep[p. 33 ff.]{book_Paolo} for all data points above $1 \times 10^{18}$\,eV in the SU(2)$_{\rm CMB}$ model, we find the best fit parameters shown in
Table\,\ref{table:BestFitSU2} and plotted in Fig.\,\ref{fig:HeinzeBestSU2} as full lines. 
For this fit, the proton excess of SU(2)$_{\rm CMB}$ below the ankle is reduced and the main contributing factor is the shallower source evolution ($m = 2.7$) in contrast to the stronger evolution $m = 4.2$ for U(1) in Heinze's best fit. The injected chemical composition and the spectral index are only mildly changed, which suggests that the shallower source evolution is enough to compensate for the increased proton horizon and the pileup below the ankle. 
Note that the proton fraction below the ankle is still too high, in disagreement with the chemical composition inferred from the $X_{\rm max}$ data (see also appendix B, Fig.\,\ref{fig:best_fit} c). Below the ankle, an additional galactic component is expected with a heavier composition.

\begin{table}[ht]
\centering
\begin{tabu}{ | l | c |[1.5pt] c | c | c |}
\hline                  
 EBL & Gilmore et al. & Element & $f_A$ \% & $I_A$ \%  \\ \hline
 models & TALYS \& Sibyll 2.3c& H & 0.001 & 0.0\\  \hline
 redshifts & $1\,-\,0$ & He & 82.3& 9.74\\  \hline
 $\gamma$ & $-0.89$ & Ni & 17.4 &  76.9\\\hline
 $R_{\textrm{max}}$ & $1.74 \times 10^{18}$\,V & Si & 0.35&11.6\\\hline
 $m$ & 2.7 & Fe & 0.01&1.72\\\hline
\end{tabu}
\protect\caption{Best fit gradient descent parameters for the SU(2)$_{\rm CMB}$ model.}
\label{table:BestFitSU2}
\end{table}

\begin{figure}[tbh]
\centering
\protect\caption{The effect of seven modified CMB photon densities on the total cosmic ray flux is shown in comparison to the normal U(1) temperature redshift relation, as obtained in \protect\cite{Heinze:2019jou} (navy blue, dashed) on top of Auger data from 2017.
The best fit parameters of the gradient descent method are used, compare Table\,\ref{table:BestFitSU2}.
The total CR flux for an SU(2) $T(z)$ relation is shown in navy blue. 
0.5 $\times$ U(1) is shown in red dotted lines, 0.75 $\times$ U(1) orange dashed
, 1.25 $\times$ U(1) yellow dot-dashed
, 1.5 $\times$ U(1) green dotted
, 1.75 $\times$ U(1) light blue dashed
, 2 $\times$ U(1) purple dot-dashed.
}
\includegraphics[width=\columnwidth]
{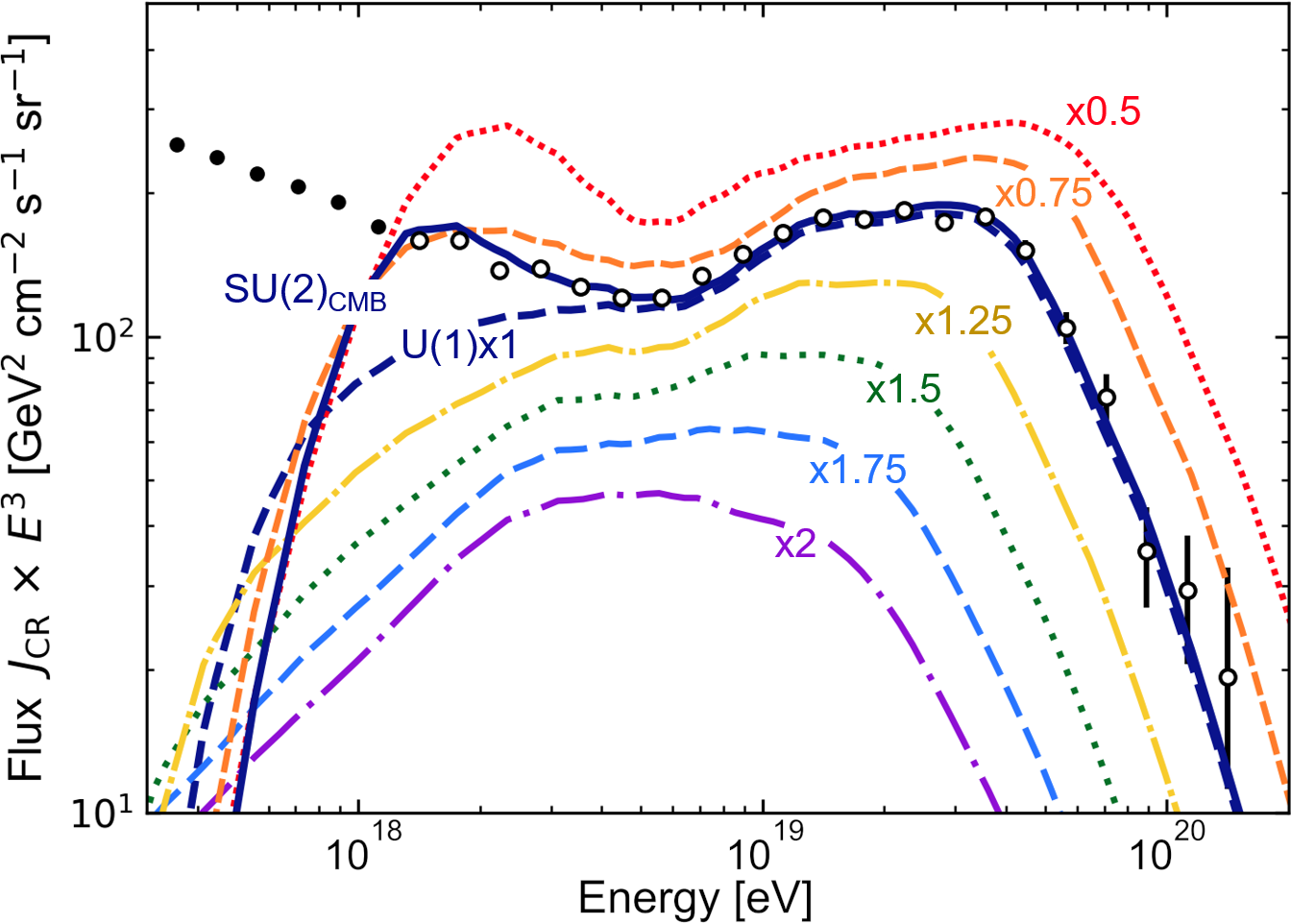}
\label{fig:HeinzeBestSU2MultipleTz}
\end{figure}

To better illustrate the SU(2) impact on UHECR propagation, Fig.\,\ref{fig:HeinzeBestSU2MultipleTz} contrasts the cosmic ray fluxes resulting with a conventional U(1) propagation employing the best fit parameters from Table\,\ref{table:BestFitSU2} and scaling the CMB photon density by different factors as shown in the curve labels.
The red dotted line in Fig.\,\ref{fig:HeinzeBestSU2MultipleTz} corresponds to SU(2)$_{\rm L}$, where CMB photons interact only with half of the UHECRs due to their handedness.

The excess in proton flux below the ankle is correlated with the CMB photon density, because these protons come from the disintegration of nuclei. However, this relation is dependent on the injection spectral index, and it is hard to distinguish an increased proton flux from an additional UHECR source and source evolution. 
Detailed directional studies which also consider the effects of magnetic fields as well as a better understanding of the chemical composition below the ankle are necessary in order to favour or disfavour the correlation between the slope of the UHECR flux below the ankle and $T(z)$. 
Note also that only hard spectra, i.e.\ $\gamma \le 0$  can increase significantly the UHECR flux below the ankle, because of the larger contribution of the highest energies in secondary protons. Soft injection spectra, e.g.\ $\gamma \approx 2$ as expected by shock acceleration, do not significantly increase the UHECR flux under SU(2)$_{\rm CMB}$.

%####################################################
%       Cosmogenic Neutrinos
%####################################################
\section{Cosmogenic Neutrinos}\label{Cosmogenic Neutrinos Model Comparisonn}

\begin{figure}[tbh]
\centering
\protect\caption{
The cosmogenic neutrino flux obtained from the gradient descent fit, Table\,\ref{table:BestFitSU2}. SU(2)$_\textrm{CMB}$ is shown in navy blue, normal $\Lambda$CDM with the corresponding cosmological parameters and U(1) photon propagation is shown in a navy blue dashed line.
The pink shaded area represents the projected sensitivity for the IceCube Gen2 radio upgrade after 5 years of observation, compare Fig.\,5 in \protect\cite{IceCube:2019pna}.
The lavender dotted line indicates the expected sensitivity for Grand200k after 3 years \protect\citep{GRAND:2018iaj}.
The dark purple and green dashed lines show 90\% CL limits from the IceCube and Pierre Auger Collaboration, respectively \protect\citep{IceCube:2018fhm,PierreAuger:2019ens}.
}
\includegraphics[width=\columnwidth]
%{pictures/gradient_descent_neutrios.png}
{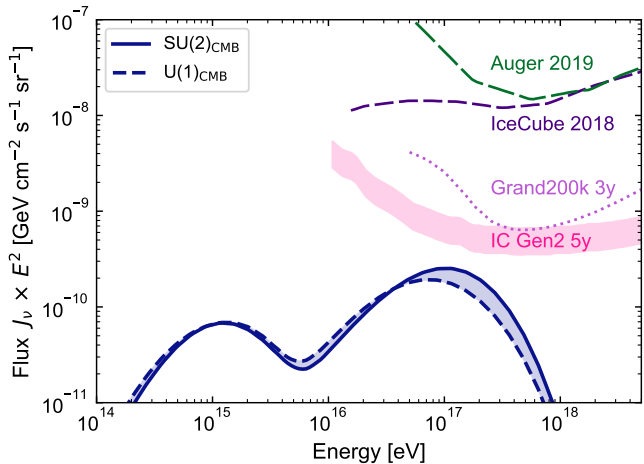}
\label{fig:cosmogenic_neutrino_fluxes}
\end{figure}

The expected cosmogenic neutrino fluxes are shown in Fig.\,\ref{fig:cosmogenic_neutrino_fluxes} for the modified temperature redshift relation under SU(2)$_{\rm CMB}$ and the normal $T(z)$ for the best fit values from the gradient descent method, Table\,\ref{table:BestFitSU2}. The neutrino fluxes for SU(2)$_{\rm CMB}$ peak at slightly higher energies and are slightly increased. The former feature is a consequence of the changed redshift dependence, which increases the energy of the GZK limit in SU(2)$_{\rm CMB}$ compared to U(1). The latter effect results from the increase in the propagation horizon of the source protons.

Figure \ref{fig:cosmogenic_neutrino_fluxes} shows that changes in the $T(z)$ relation of the CMB only affect the cosmogenic neutrino flux for energies around $10^{17}$\,eV. The peak at around $10^{15}$\,eV, stemming mostly from the decay of neutrons from photodisintegration (see e.g.\ \cite{Ave:2004uj}) is mostly unaffected except for being slightly narrower due to reduced pair production losses.\\

In addition to the cosmogenic neutrinos, the photopion production with the CMB also generates $\gamma$-rays and the resulting flux at Earth in the case of a SU(2)$_{\rm CMB}$ would be slightly enhanced compared to the U(1)$_{\rm CMB}$  due to the increased horizon in the absence of $\gamma\gamma$-pair production. However, $\gamma\gamma$-pair production and inverse Compton scattering with the EBL are the dominant interactions, in particular for $\gamma$-ray energies $\lesssim 100\,\text{TeV}$, therefore, after the cascading of photon we expect no significant difference in the cosmogenic $\gamma$-ray flux between the SU(2)$_{\rm CMB}$ and the U(1)$_{\rm CMB}$.

%####################################################
%                   Summary and Outlook
%####################################################
\section{Summary and Outlook}\label{Summary and Outlook}

In this paper, we examined the impact of locally non-linear modification of the CMB temperature redshift relation $T(z)$ on the fit to ultra-high energy cosmic rays and the corresponding cosmogenic neutrinos. 

The reduction of the CMB densities is found to affect significantly the interaction lengths of UHECRs with CMB photons in the redshift range of relevance for UHECR propagation, resulting in extended horizons for protons and UHECR nuclei. However, the increase in interaction lengths has only a modest effect on the observed UHECR flux due to interactions with the EBL, which then become dominant for the energies of relevance. Hence, a comparison to an existing fit of UHECRs yields similar flux of UHECRs nuclei but differs considerably for protons where a pronounced bump appears below the ankle for the SU(2)$_{\rm CMB}$ for hard injection spectra.\\

In order not to exceed the total UHECR flux and to agree with Auger data in the case of a hard injection spectrum, a shallower source evolution of cosmic ray sources of $m \approx 2.7$ is needed, which is more in line with SBGs and GRBs than with AGNs. 
This is in agreement with recent studies that consider arrival directions and extragalactic magnetic fields for energies beyond the ankle ($\ge 5 \times 10^{18}$\,eV) \citep{Bister:2023lyb}.\\

While the confirmation of the SU(2)$_{\rm CMB}$ description requires further studies, the present work provides constraints for its validity. The independent determination of the redshift evolution of UHECR sources has the potential to reject the SU(2)$_{\rm CMB}$ temperature redshift relation for {\it hard} injection spectra: for a steeper cosmic ray source evolution, the predicted proton contribution below the ankle would be in tension with observations. \\

Since there is currently no firm preference for a specific UHECR source class \citep{PierreAuger:2022axr}, we would like to add modified $T(z)$ and in particular in the case of SU(2)$_{\rm CMB}$ to the discussion. 
This adds another tool to discriminate potential source classes and vice versa, constraining the sources by other means while simultaneously improving the knowledge of the UHECR composition may lead to a direct probe of $T(z)$ of the CMB in the future.\\
\vspace{-3mm}
\section{Data availability}
\noindent The simulation scripts used in this study are available upon request, and the authors welcome inquiries for collaboration.
% I would like to add a GitLab repository here

\section{Acknowledgements}
JM acknowledges insightful discussions with Ralf Hofmann and Wolfgang Rhode.\\

This work is supported by SFB 1491 (Project A3)
and partially by the Vector Foundation under grant number P2021-0102. 
LM's work is supported by the DFG under grant number 445990517 (KA 710).
%%%%%%%%%%%%%%%%% APPENDICES %%%%%%%%%%%%%%%%%%%%%

\bibliographystyle{mnras}
\bibliography{2023_SU2_Bib_template}%BibSU2CMB

\clearpage
\newpage

%################################# A ###################################

\section*{Appendix A: Best fit Heinze \texorpdfstring{SU(2)$_{\rm CMB}$}{SU(2)CMB}}\label{Appendix A}

\setcounter{figure}{8}
\setcounter{table}{2}

\begin{samepage}

The best fit from \cite{Heinze:2019jou} is reproduced in Fig.\,\ref{fig:best_fit_Heinze} a) alongside the cosmogenic neutrino flux, Fig.\,\ref{fig:best_fit_Heinze} b), the $X_{\rm max}$ data Fig.\,\ref{fig:best_fit_Heinze} c) and $\sigma(X_{\mathrm{max}})$ data, Fig.\,\ref{fig:best_fit_Heinze} d) overlaid by three hadronic interaction models.\\

All $\chi^2$ for the $X_{\mathrm{max}}$ of the best fit in Fig.\,\ref{fig:best_fit_Heinze} c) and the $\chi^2$ for $\sigma(X_{\mathrm{max}})$ of the best fit in Fig.\,\ref{fig:best_fit_Heinze} d) are shown in table\,\ref{table:BestFitSU2_Chi2_of_Xmax_relative_Heinze}.

\begin{figure}[H]
\centering
\onecolumn
\protect\caption{
\textbf{a)} Spectral fit to the 2017 Auger spectral flux data 
from the best fit parameters in \protect\cite{Heinze:2019jou}, see Tab.\,\ref{table:HeinzeTable}. \protect\cite{Heinze:2019jou} assume a normal U(1) temperature redshift relation (dashed lines, total flux dashed navy blue). The total cosmic ray flux with an 
SU(2) temperature redshift relation is shown with a blue solid line. The $\chi^2$ only consider data points including the ankle region (white dots), according to Heinze's choice.
\textbf{b)} The cosmogenic neutrino flux obtained from the fit in (a) in SU(2)$_\textrm{CMB}$ is shown in navy blue solid, \protect\cite{Heinze:2019jou} in navy blue dashed lines.
The pink shaded area represents the projected sensitivity for the IceCube Gen2 radio upgrade after 5 years of observation, compare Fig. 5 in \protect\cite{IceCube:2019pna}.
The lavender doted line indicates the expected sensitivity for Grand200k after 3 years \protect\cite{GRAND:2018iaj}.
The dark purple dashed line shows 90\% CL limits from the IceCube Collaboration (2018) \protect\citep{IceCube:2018fhm}.
And the green dashes line represents the 90\% CL limit from the 
Pierre Auger Collaboration (2019) \protect\citep{PierreAuger:2019ens}.\\
\textbf{c)} The Auger 2017 $\langle$X$_\textrm{max}\rangle$ and 
\textbf{d)} $\sigma$(X$_\textrm{max}$) data \protect\citep{Bellido:2017cgf}, on top of three different air-shower model expectations: 
Epos-LHC  \protect\citep{Pierog:2013ria} (solid lines), 
Sibyll 2.3c \protect\citep{Riehn:2015oba} (dashed bold lines) and 
QGSJET-II.04 \protect\citep{Ostapchenko:2010vb} (dotted lines). For calculating the relative $\chi^2$ the same energy ranges as in Fig.\,\ref{fig:best_fit} c) and d) have been chosen for better compatibility.
}
\twocolumn
\includegraphics[width=18cm]{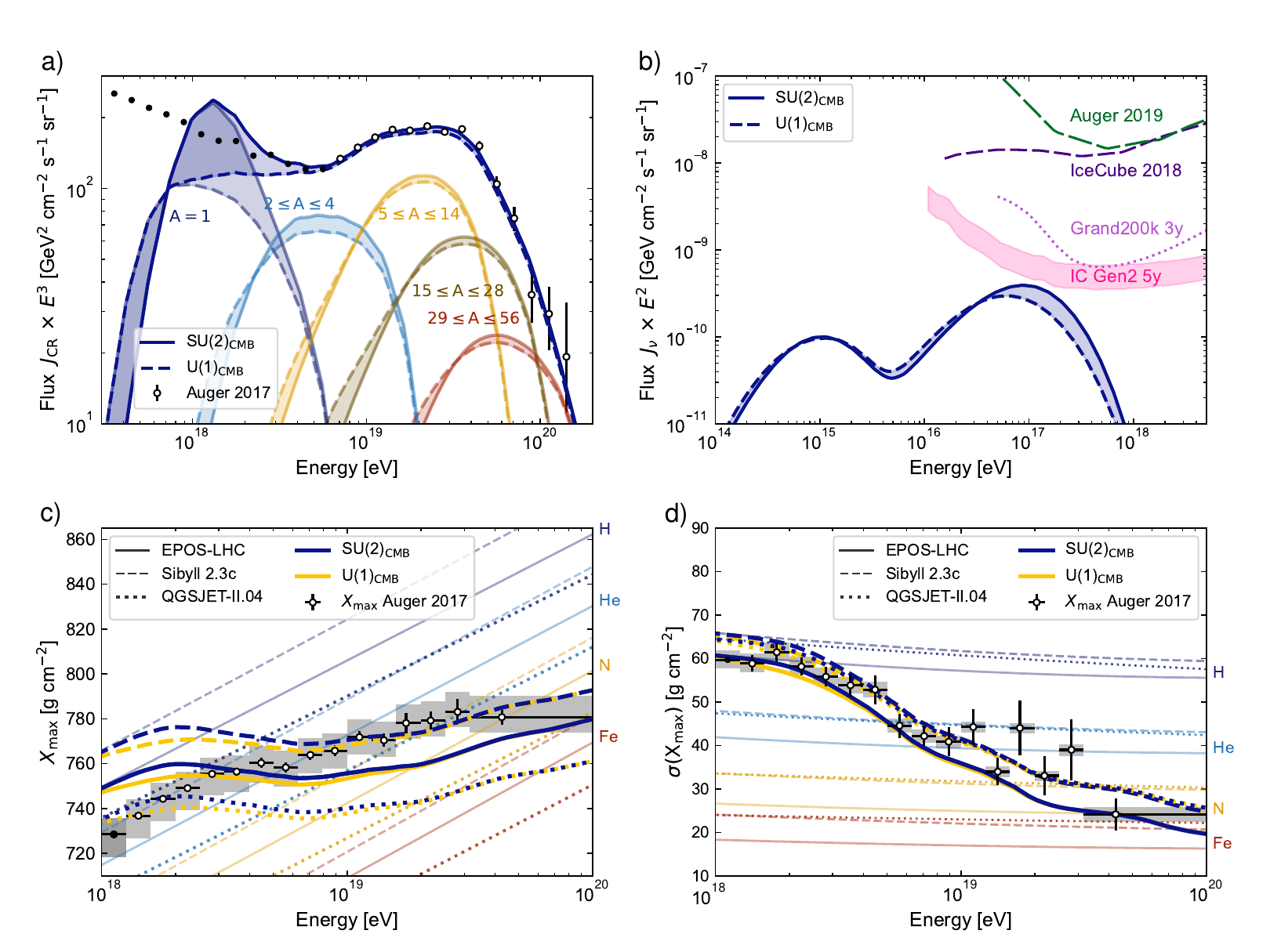}
%{pictures/Heinze_panel.png}
\label{fig:best_fit_Heinze}
\end{figure}

\newpage

\begin{table}[H]
\centering
\begin{tabu}{ | l | c | c | c | c | }
\hline
reduced $\chi^2$ for &
 \multicolumn{2}{c|}{
$X_{\mathrm{max}}$ in Fig.\,\protect\ref{fig:best_fit_Heinze} c) }& 
 \multicolumn{2}{c|}{
$\sigma(X_{\mathrm{max}})$ in Fig.\,\protect\ref{fig:best_fit_Heinze} d) }\\ \hline
 Model & U(1) & SU(2)$_{\mathrm{CMB}}$ & U(1)& SU(2)$_{\mathrm{CMB}}$\\ \hline
 %EPOS-LHC & 1.49 & 1.39 & 10.96 & 10.96  \\  \hline
 EPOS-LHC & 2.21 & 2.2 & 11.76 & 10.96  \\  \hline
 Sibyll 2.3 c & 0.91 & 1.36 & 4.02 & 4.25 \\  \hline
 QGSJET-II.04 & 5.39 & 4.79 & 3.64 & 3.75\\  \hline
\end{tabu}
\protect\caption{The reduced $\chi^2$ for the $\sigma(X_{\mathrm{max}})$ of the best fit in Fig.\,\protect\ref{fig:best_fit} d). Seven degrees of freedom were assumed.
}
\label{table:BestFitSU2_Chi2_of_Xmax_relative_Heinze}
\end{table}

\end{samepage}
\pagebreak
\clearpage
%#######################################################################
%################################# B ###################################

\section*{Appendix B: Best fit \texorpdfstring{SU(2)$_{\rm CMB}$}{SU(2)CMB}}\label{Appendix B}

\begin{samepage}

In order to mitigate the proton excess below the ankle in the reproduced model from \cite{Heinze:2019jou}, a descending algorithm is used. 
It minimizes the $\chi^2$ for the white spectral data points in Fig.\,\ref{fig:best_fit} a), $X_{\rm max}$ in Fig.\,\ref{fig:best_fit} c) and $\sigma(X_{\mathrm{max}})$ in Fig.\,\ref{fig:best_fit} d).
The best fit parameter to the Pierre Auger Collaboration data from 2017 are shown in Table\,\ref{table:BestFitSU2}.\\

All $\chi^2$ for the $X_{\mathrm{max}}$ of the best fit in Fig.\,\ref{fig:best_fit} c) and the $\chi^2$ for $\sigma(X_{\mathrm{max}})$ of the fit in Fig.\,\ref{fig:best_fit} d) are given in table\,\ref{table:BestFitSU2_Chi2_of_Xmax_relative}.

\begin{figure}[tbh]
\centering
\onecolumn
\includegraphics[width=18cm]{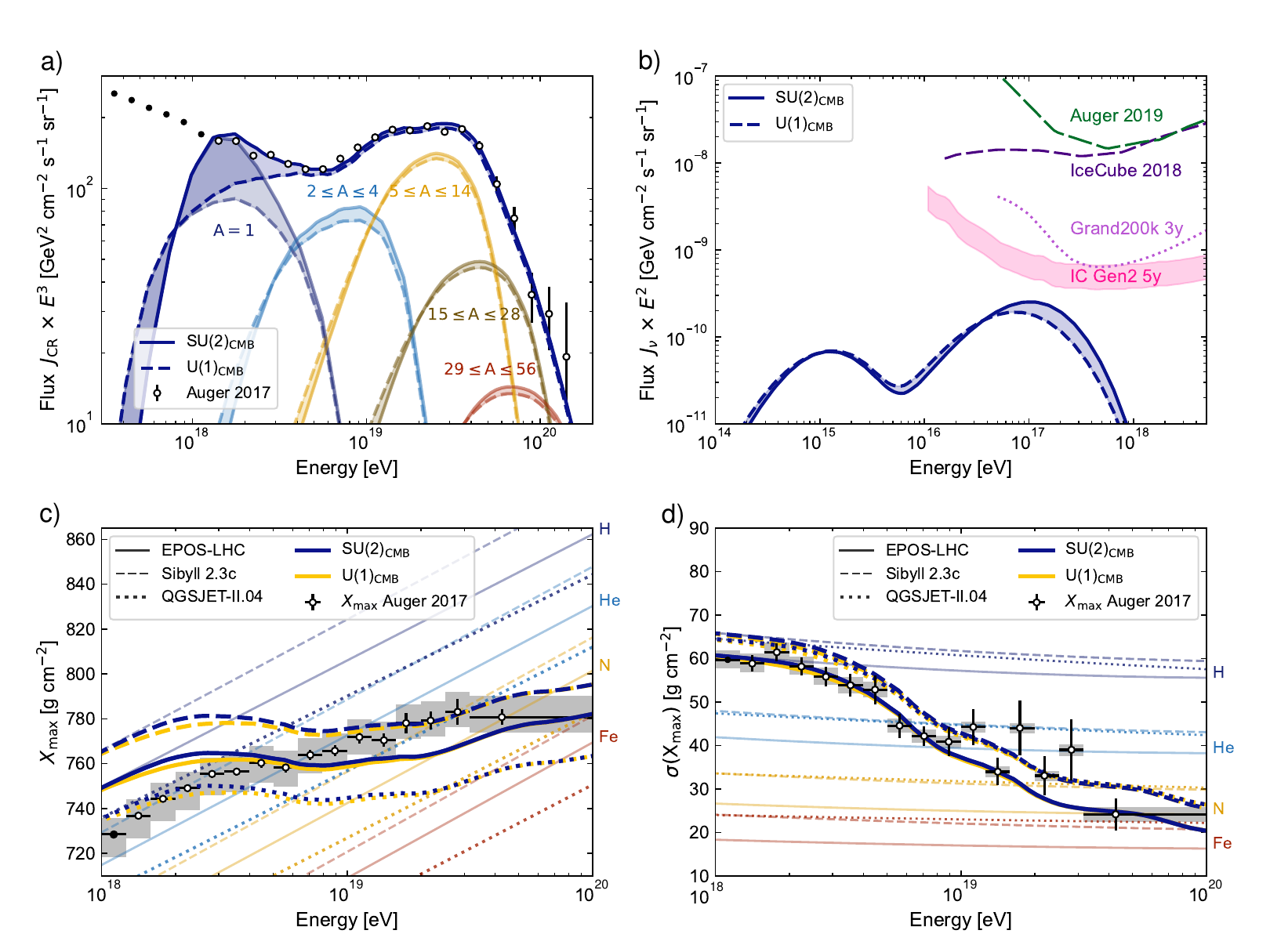}
\caption[]{
\textbf{a)} Spectral fit to the 2017 Auger spectral flux data 
from the best fit parameters in \cite{Heinze:2019jou}, see Tab.\,\ref{table:HeinzeTable}. \cite{Heinze:2019jou} assume a normal U(1) temperature redshift relation (dashed lines, total flux dashed navy blue). The total cosmic ray flux with an 
SU(2) temperature redshift relation is shown with a blue solid line. The $\chi^2$ only consider data points including the ankle region (white dots), according to Heinze's choice.
\textbf{b)} The cosmogenic neutrino flux obtained from the fit in (a) in SU(2)$_\textrm{CMB}$ is shown in navy blue solid, \citep{Heinze:2019jou} in navy blue dashed lines.
The pink shaded area represents the projected sensitivity for the IceCube Gen2 radio upgrade after 5 years of observation, compare Fig.\ 5 in \cite{IceCube:2019pna}.
The lavender doted line indicates the expected sensitivity for Grand200k after 3 years \cite{GRAND:2018iaj}.
The dark purple dashed line shows 90\% CL limits from the IceCube Collaboration (2018) \citep{IceCube:2018fhm},
and the green dashes line represents the 90\% CL limit from the 
Pierre Auger Collaboration (2019) \citep{PierreAuger:2019ens}.\\
\textbf{c)} The Auger 2017 $\langle$X$_\textrm{max}\rangle$ and 
\textbf{d)} $\sigma$(X$_\textrm{max}$) data \citep{Bellido:2017cgf}, on top of three different air-shower model expectations: 
Epos-LHC \citep{Pierog:2013ria} (solid lines), 
 Sibyll 2.3c \citep{Riehn:2015oba} (dashed bold lines) and 
 QGSJET-II.04 \citep{Ostapchenko:2010vb} (dotted lines).
}
\label{fig:best_fit}
\twocolumn
\end{figure}

\newpage

\begin{table}[H]
\centering
\begin{tabular}{ | l | c | c | c | c | }
\hline
reduced $\chi^2$ for & \multicolumn{2}{c|}{$X_{\mathrm{max}}$ in Fig.\,\protect\ref{fig:best_fit} c)}  & 
\multicolumn{2}{c|}{$\sigma(X_{\mathrm{max}})$ in Fig.\,\protect\ref{fig:best_fit} d)} \\ \hline
 Model & U(1) & SU(2)$_{\mathrm{CMB}}$ & U(1) & SU(2)$_{\mathrm{CMB}}$\\ \hline
 EPOS-LHC & 0.98 & 1.01 & 8.97 & 8.53 \\  \hline
 Sibyll 2.3 c & 1.57 & 2.0 & 4.86 & 5.18\\  \hline
 QGSJET-II.04 & 3.76 & 3.45 & 4.11 & 4.36\\  \hline
\end{tabular}
\protect\caption{The reduced $\chi^2$ for the $\sigma(X_{\mathrm{max}})$ of the fit in Fig\,\protect\ref{fig:best_fit} d). Seven degrees of freedom were assumed.
}
\label{table:BestFitSU2_Chi2_of_Xmax_relative}
\end{table}

\end{samepage}
\pagebreak
\clearpage

\label{lastpage}
\end{document}